\documentclass[letterpaper]{article}
\usepackage{aaai16}
\usepackage{array}
\usepackage{balance}
\usepackage{blindtext}
\usepackage{color}
\usepackage{courier}
\usepackage{multirow}
\usepackage{enumitem}
\usepackage{graphics}
\usepackage{helvet}
\usepackage{graphicx}
\usepackage{subfig}
\usepackage{epstopdf}
\usepackage{rotating}
\usepackage{times}
\usepackage{todonotes}
\usepackage{txfonts}
\usepackage{url} 

\newcolumntype{L}[1]{>{\raggedright\let\newline\\\arraybackslash\hspace{0pt}}m{#1}}
\newcolumntype{C}[1]{>{\centering\let\newline\\\arraybackslash\hspace{0pt}}m{#1}}
\newcolumntype{R}[1]{>{\raggedleft\let\newline\\\arraybackslash\hspace{0pt}}m{#1}}

\begin{document}

\title{The Emotional and Chromatic Layers of Urban Smells}

\author{Daniele Quercia\\Bell Labs\\\textit{daniele.quercia@gmail.com} \And
Luca Maria Aiello\\Yahoo Labs\\\textit{alucca@yahoo-inc.com} \And
Rossano Schifanella\\University of Turin\\\textit{schifane@di.unito.it}}
\maketitle

\begin{abstract}
\begin{quote}
People are able to detect up to 1 trillion odors. Yet, city planning is concerned only with a few bad odors, mainly because odors are currently captured only through complaints made by urban dwellers.  To capture both good and bad odors, we resort to a methodology that has been recently proposed and relies on tagging information of geo-referenced pictures. In doing so for the cities of London and Barcelona, this work makes three new contributions. We study 1) how the urban smellscape changes in time and space; 2) which emotions people share at places with specific smells; and 3) what is the color of a smell, if it exists. Without social media data, insights about those three aspects have been difficult to produce in the past, further delaying the creation of urban restorative experiences.
\end{quote}
\end{abstract}

\section{Introduction} \label{sec:introduction}

The intensity of modern city life calls for urban places that offer relief  (e.g., greenery), and one way of designing such places (designing what architects call ``restorative environmental experiences'') is to map sensory perceptions at city scale.

Traditional research on the sensory urban experience has primarily focused on the visual dimension. Back in the 60s, Kevin Lynch, for example, showed how what we are able to see and remember of a city contribute to our ability to navigate it and, ultimately, to our well-being as residents~\cite{lynch1960}. On the other hand, previous research on olfactory and sonic perceptions has mainly explored negative characteristics~\cite{victoria2013,schafer1993soundscape}:  urban sounds have been equated to noise, and smells to nuisance odors, so much so that both end up being ``guilty until proven innocent''~\cite{fox06smell}.  

To see why, take smell. A comprehensive study of it is made difficult by the simple fact that odors are hard to capture (Section~\ref{sec:related}). To partly fix that, we have recently proposed a methodology with which social media data can be used to capture the smellscape of entire cities~\cite{quercia15smelly}. After creating a dictionary of smell-related words, we matched those words with social media metadata (e.g., Flickr and Instagram tags). By building upon that methodology (Section~\ref{sec:background}), we recreate the smellscapes of London and Barcelona. We capitalize on that mapping to make three new contributions:

\begin{itemize}[leftmargin=*]
\item We study the temporal and spatial dynamics of smell (Section~\ref{sec:time}). As for temporal aspects, we find that the most seasonal smells are those  of plant and nature, and the most olfactory pleasant months are in Spring. As for spatial aspects, specific areas in the city turn out to be characterized by distinctive smells (e.g., parks), while non-central areas tend to offer bland smellscapes.
\item We study the  relationship between smell and emotions (Section~\ref{sec:emotions}). Streets with  smells of nature and food are associated with positive emotion words (picture tags), while those with smells for metro and waste are associated with negative ones. 
\item We conclude this work by studying the relationship between smell and color (Section~\ref{sec:colors}). We do so by testing whether a specific color is predominantly present in pictures associated with a given smell. We find that certain smells are associated with definitive colors (e.g.,  traffic is mainly black), while others span a variety of colors (e.g., food takes multiple colors).
\end{itemize}

\section{Related work}
\label{sec:related}

Since  we are interested in capturing smell words from picture tags on social media, we look at what computer scientists have done in the area of text mining on social media. 

\mbox{ } \\
\textbf{Use of language across time.}  Dodds \emph{et al.} analyzed text on tweets to remotely sense societal-scale levels of happiness and studied how word use changed as a function of time~\cite{temporal11dodds}. They collected tweets posted by over 63 million users in 33 months, and built a tunable hedonometer that analyzed word usage in real-time to gauge ``happiness'' levels. They found fascinating temporal variations in happiness over timescales ranging from hours to years.  At year level, they found that, after an upward trend starting from January to April 2009, average happiness gradually decreased. 
At month level, they saw average happiness gradually increased towards the end of each year. At week level, they found peaks over the weekends, and nadirs on Mondays and Tuesdays. By looking at individual days, they were able to show that 
negative days were seen during unexpected societal trauma such as the 2008 Bailout of the US financial system and the February 2010 Chilean earthquake. A year later, Lansdall-Welfare \emph{et al.} analyzed 484 million tweets generated by more than 9.8 million users from the United Kingdom over 31 months. They did so to  study the impact of the economic downturn and  social tensions on the use of language on Twitter~\cite{lansdall2012}. More specifically, they studied the use of emotion words classified in four categories taken from the tool `WordNet Affect': anger, fear, joy and sadness. This resulted into  146 anger words, 92 fear words, 224 joy words and 115 sadness words. The authors found that periodic events such as Christmas, Valentine and Halloween were associated with very similar use of words, year after year. On the other hand, they observed two main negative change-points, one  occurring in October 2010, when the government announced cuts to public spending; and the other in Summer 2011, when riots broke out in various UK cities. Interestingly, the increase use of negative emotion words  preceded, not followed, those events, suggesting predictive ability. Golder and Macy studied the 500 million English tweets that  2.4 million users  produced during almost 2 years. Based on their hour-by-hour analysis, they found that offline patterns of mood variations also hold on Twitter: mood variations were associated with  seasonal changes in day length. People also changed their mood as the working day progressed and were happier during weekends~\cite{Golder2011}.

\mbox{ } \\
\textbf{Use of language across space.} Schwartz \emph{et al.} tested whether the language used in tweets is predictive of the subjective well-being of people living in US counties~\cite{characterizing2013}. They did so by collecting  a random sample of tweets in 1,293 US counties. By correlating the word use with subjective well-being  as measured by representative surveys, they found positive correlations with pro-social activities, exercise, and engagement with personal and work life, and negative correlations with  words of disengagement. Those findings were in line with existing well-being studies in the social sciences. Frank \emph{et al.} went on characterizing the mobility patterns of 180,000 individuals~\cite{frank2013}. In so doing, the researchers were able to characterize changes in the use of language in relation to people's mobility. They found that tweets written close to a user's center of mass (typical location) are slightly happier than those written 1 km away, which is the distance representative of a short daily commute to work. Beyond this least happy distance, they found that the use of positive emotion words increased logarithmically with distance. This pattern did hold when they moved to study a user's radius of movement. The larger a user's radius, the more happier words the user tended to use. Finally, De Choudhury \emph{et al.} studied whether people exposed to chronic violence lowered affective responses in their Twitter posts~\cite{DeChoudhury14narco}. To this end, they collected all of the Spanish tweets that were mentioning one of the four Mexican cities of Monterrey, Reynosa, Saltillo, and Veracruz. Between Aug 2010 and Dec 2012, these four cities were affected by  protracted violence in the context of the ``Mexican Drug War''. The researchers found that, while violence was on the rise offline, negative emotional expression online was declining and emotional arousal and dominance were raising: both aspects are known psychological correlates  of population desensitization. This suggests that chronic exposure to violence is indeed associated with signs of desensitization in social media postings.

\mbox{ } \\
The focus of our work is on \textit{urban} smell, so summarizing what computer scientists and urban planners have already done in the area is in order. 

\mbox{ } \\
\textbf{Smell in Computer Science.} In 2004, Kaye showed that the vast majority of work in Computer Science and, more specifically, in Human-Computer Interaction ``involves our senses of sight and hearing, with occasional forays into touch.''~\cite{kaye04}  Since then, work in HCI has focused on smell technologies that are able to \emph{capture} and  \emph{generate} smells. Bodnar and Corbett proposed smell-based notifications and showed they were less disruptive than visual notifications or auditory ones~\cite{bodnar04}. A couple of years later, Brewster \emph{et al.} designed Olfoto, a photo tagging tool in which smell was used to search the collection~\cite{brewster06}. More recently, to go beyond capturing smell, researchers have explored the idea of transmitting it as well,  and they did so over the Internet~\cite{ranasinghe11}. More generally, experiences with smell have profound and multi-faceted implications for technology~\cite{Obrist14}.

\mbox{ } \\
\textbf{Smell in Urban Planning.} People are able to detect up to 1 trillion smells~\cite{bushdid14}, yet city planning focuses on a few bad smells only. City officials entirely rely on complaints to capture smell and ultimately inform urban planning. In the research world, smell has been recorded in a variety of ways~\cite{victoria2013}: with `nose trumpets' that capture four main olfactory aspects (i.e., odor character, odor intensity, and duration); with web maps that elicit smell words from users in a crowdsourcing way; and with sensory walks~\cite{diaconu2011senses} in which groups of people are asked to walk around the city and record what they smell. Unfortunately, those ways of collecting smell information require  substantial public engagement to be effective. 

\mbox{ } \\
\textbf{Missed opportunities.} Based on this literature review, one can see that, in both Computer Science and Urban Planning, there is no effective solution to capture smell (both its positive and negative aspects) at scale, and such an inability has likely limited the scholarly production in those two fields.

\section{Urban Smell from Social Media}
\label{sec:background}

\begin{figure}[t!]
\begin{center}
\includegraphics[width=0.99\columnwidth]{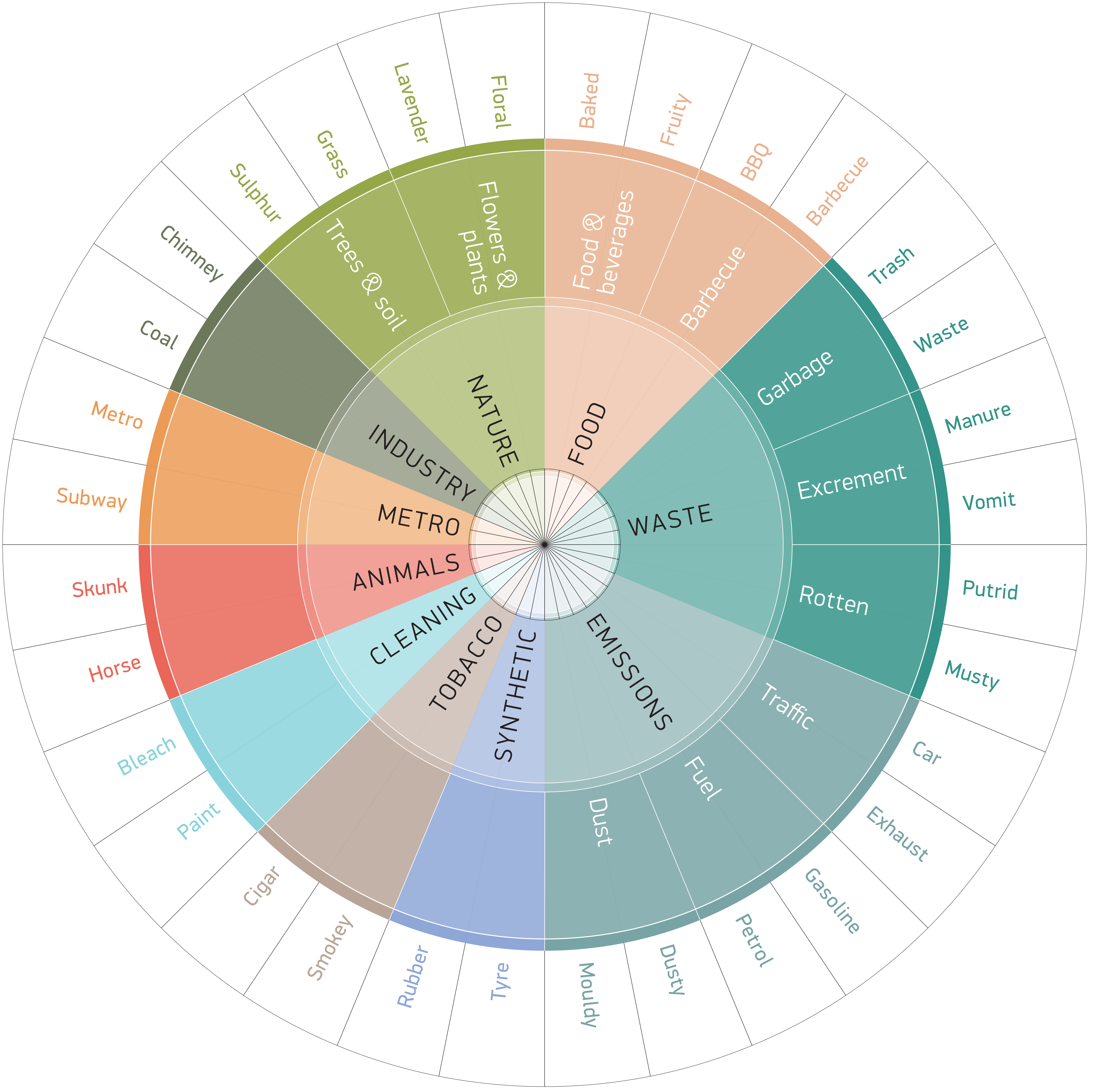}
\caption{Urban smell taxonomy. Top-level categories are in the inner circle; second-level categories, if available, are in the outer ring; and examples of words are in the outermost ring. For space limitation, in the wheel, only the first categories (those in the inner circle) are complete, while subcategories and words represent just a sample.}
\label{fig:taxonomy}
\end{center}
\end{figure}

\begin{figure*}[t]
    \subfloat[London \label{fig:london_map}]{%
     \includegraphics[width = 0.49\textwidth]{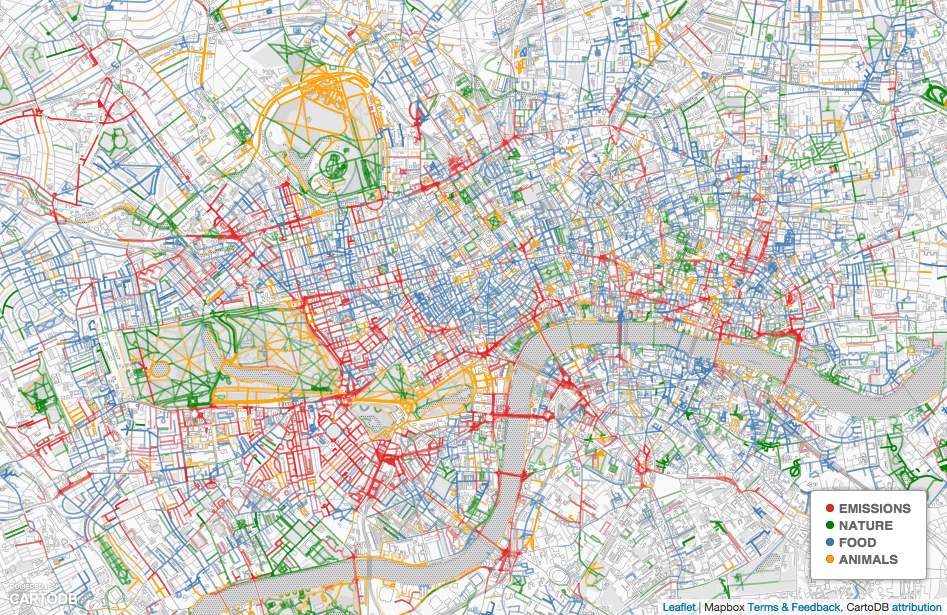}
    }
    \hfill
    \subfloat[Barcelona \label{fig:bcn_map}]{%
   \includegraphics[width = 0.49\textwidth]{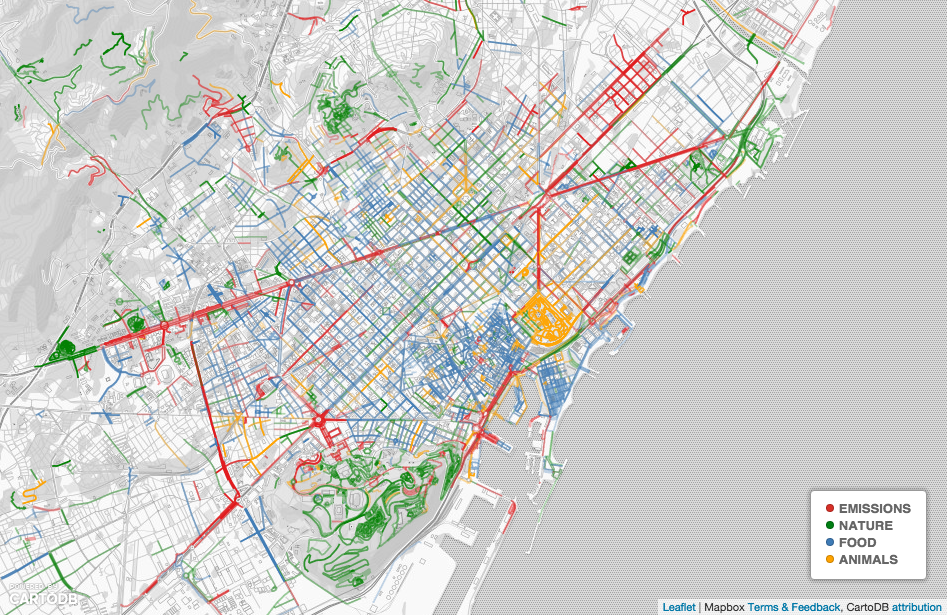}\\
    }
\caption{Urban Smell Maps for London and Barcelona. They map the fraction $f_{S@l}$ of smell category $S$ at each street segment $l$. Both cities have rich smellscapes in which odors are distributed in predictable ways. Emissions are associated with trafficked roads, nature with greenery spots, food with central parts of the cities, and animal odors with the zoos.}
\label{fig:maps_basenotes}
\end{figure*}

To partly fix that, we have recently proposed a new way of collecting odor information at scale without requiring a massive public engagement. This way simply analyzes data implicitly generated by social media users (e.g., photo tags)~\cite{quercia15smelly}, and it unfolds in three main steps:

\mbox{ } \\
\textbf{Step 1: Collecting smell-related words.} We conducted ``smellwalks'' around seven cities in UK, Europe, and USA. Locals were asked to walk around their city, identify distinct odors, and take notes (to, e.g., report smell descriptors).  As a result of those sensory walks, smell-related words were recorded and classified, resulting in the first urban smell dictionary. The smell dictionary contains words describing the smell itself (e.g., grassy), which often coincide with the object that emits the smell (e.g., musk, chocolate, pine).

\mbox{ } \\
\textbf{Step 2: Matching smell words with social media.} For the cities of Barcelona and London, we collected geo-referenced tags from 17M Flickr public photos and  436K Instagram photos, and  1.7M geo-referenced tweets from Twitter (those tweets include neither retweets nor direct replies). We then matched those tags and tweets with the words in the smell dictionary. In this work, we were able to use the same Flickr dataset, which is the biggest one and has the highest temporal coverage, spanning a time frame of more than 10 years (January 2005 to October 2015).

\mbox{ } \\
\textbf{Step 3: Organizing smell words into a dictionary.} To structure this large and apparently unrelated set of words, we built a \emph{co-occurrence graph} where nodes are smell words, and undirected edges are weighted by the number of times the two words co-occur in the same image.  Upon this graph, we found that ten ``clusters of words''  (shown in the inner circle of Figure~\ref{fig:taxonomy}) best describe the graph structure. One of those clusters is  `Emissions', for example. This cluster  has many subcategories (not all shown in the mid-layer of Figure~\ref{fig:taxonomy} for space limitation). One of those subcategories is `Fuel', which, in turn, is associated with some of the 258 smell-related word (with, e.g.,   `gasoline', `petrol'). Interestingly, this categorization (which is purely data-driven) strikingly resembles previous hand-made classifications in olfactory research~\cite{victoria2013} and has been found to have ecological validity~\cite{quercia15smelly} (i.e., the main 10 categories tend to be geographically orthogonal to each other). The dictionary is made available\footnote{\url{http://goodcitylife.org/smellymaps/}} for non-commercial purposes to anyone who wishes to translate it in a language other than the ones already at disposal. The idea is that, in the long term, dictionaries in different languages will be freely available to many stakeholders, from artists and designers to scientists. 

\begin{figure}[t]
\begin{center}
\includegraphics[width=0.99\columnwidth]{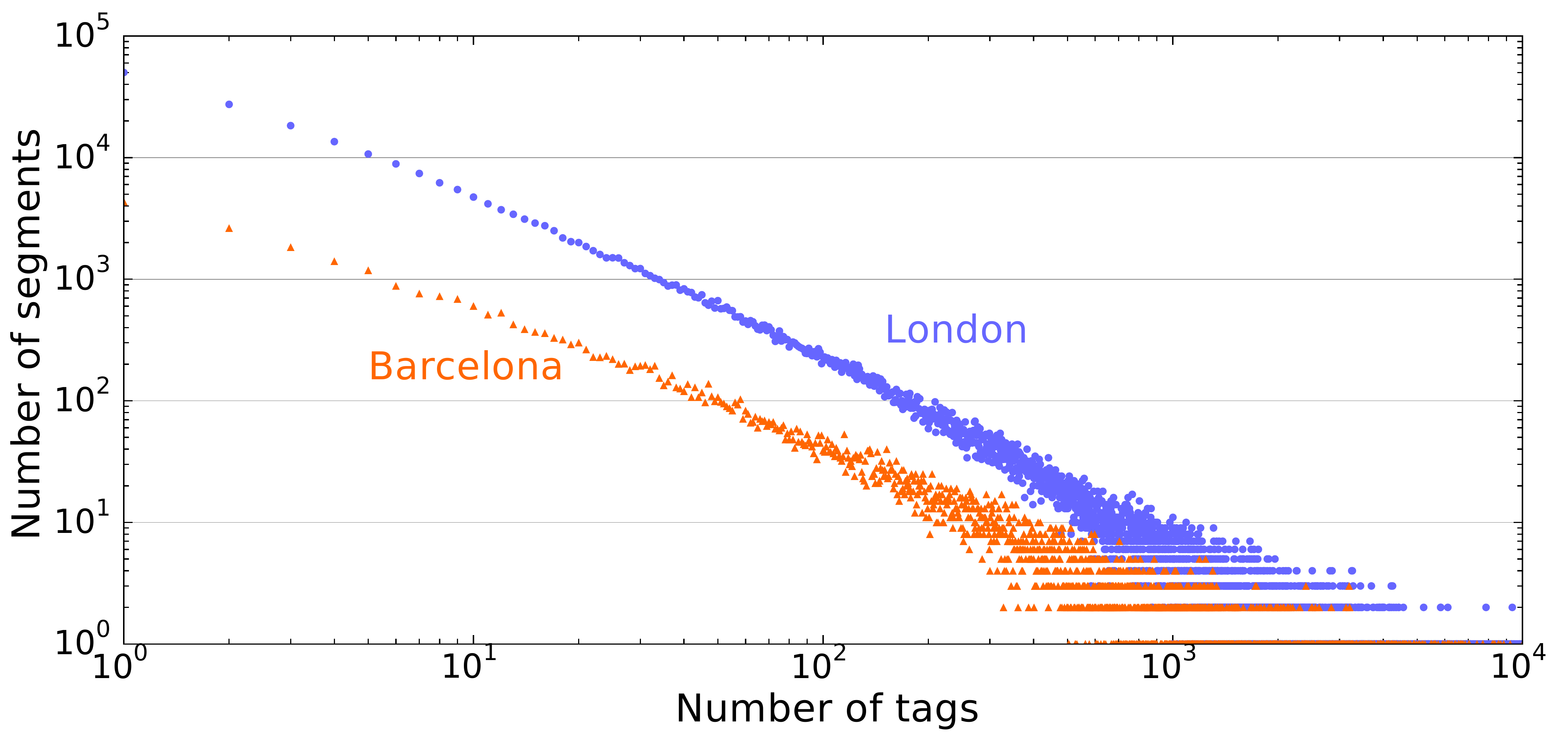}
\caption{Number of tags per street segment in London and Barcelona. Many streets have a few tags, and only a few streets have a massive number of them. London has $263K$ segments with at least one tag, Barcelona $34K$.}
\label{fig:photo_per_segment_distr}
\end{center}
\end{figure}

\begin{figure}[t!]
\begin{center}
\includegraphics[width=0.99\columnwidth]{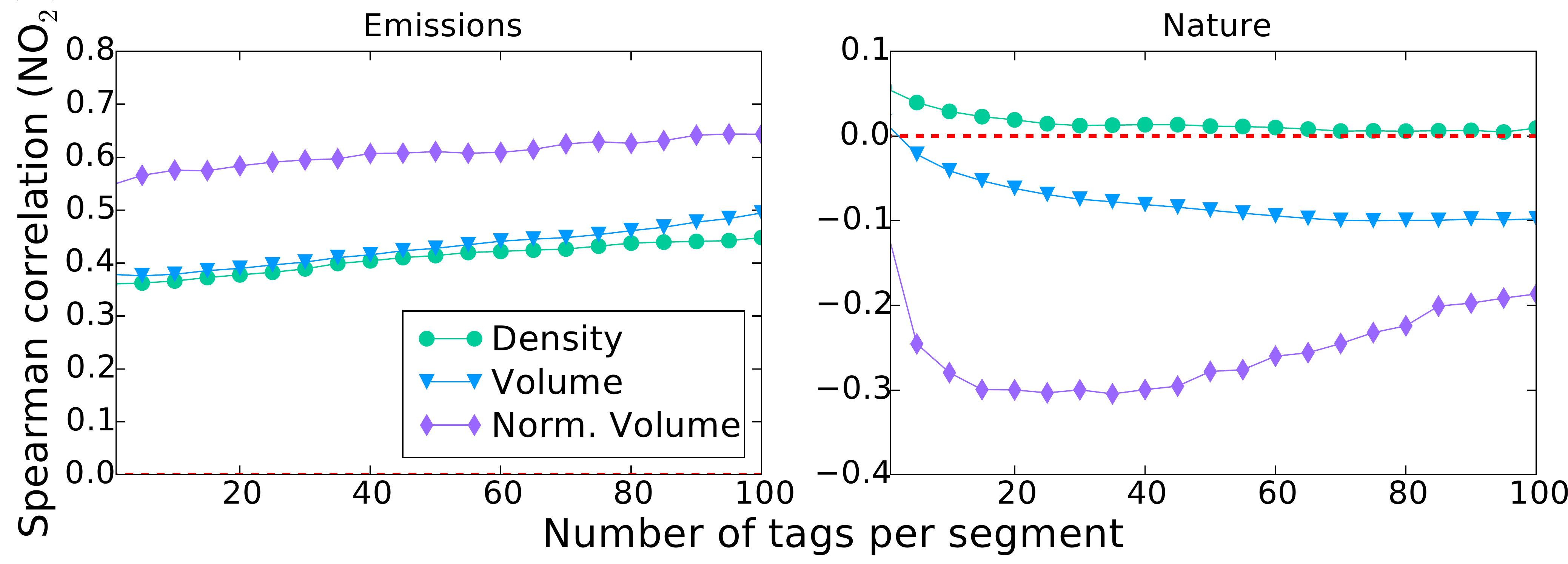}
\caption{Spearman correlation between the presence of smell categories $f_{S@l}$ and pollution levels for different ways of aggregating smell tags: density  (number of tags over a street segment's area), raw volume count, and relative volume (or fraction).}
\label{fig:aggregation_method_vsN}
\end{center}
\end{figure}

Social media data is biased, and that might make it difficult to use that data for urban olfactory analysis. To validate such a use, we collect data about the three air pollutants -- Nitrogen oxides (NO2), coarse particles (PM10), and fine particles (PM2.5) -- for both London (from King's College London API\footnote{\url{http://api.erg.kcl.ac.uk}}) and Barcelona (from the authors of~\cite{beevers13air}).  We map and use the pollution data onto $34K$ street segments in Barcelona and about $263K$ in London. A segment is a street's portion between two road intersections. Most segments have a few tags, while a few have many tags (Figure~\ref{fig:photo_per_segment_distr}). We then compute the fraction of each segment's tags that belong to a given smell category, so each segment comes with ten smell descriptors (as we have 10 smell categories). We compute and map the \emph{fraction} of tags in smell category $S$ at each location $l$  (Figure~\ref{fig:maps_basenotes}):
\begin{equation}
f_{S@l}= \frac{\textrm{\#tags in smell category } S \textrm{at location } l}{\textrm{\#tags in any smell category at location } l}
\label{eqn:fraction-smell}
\end{equation}
We map neither volume nor density because the use of the fraction yields the strongest correlation between air pollutants and smell categories (Figure~\ref{fig:aggregation_method_vsN}).

\begin{figure}[t!]
\begin{center}
\includegraphics[width=0.99\columnwidth]{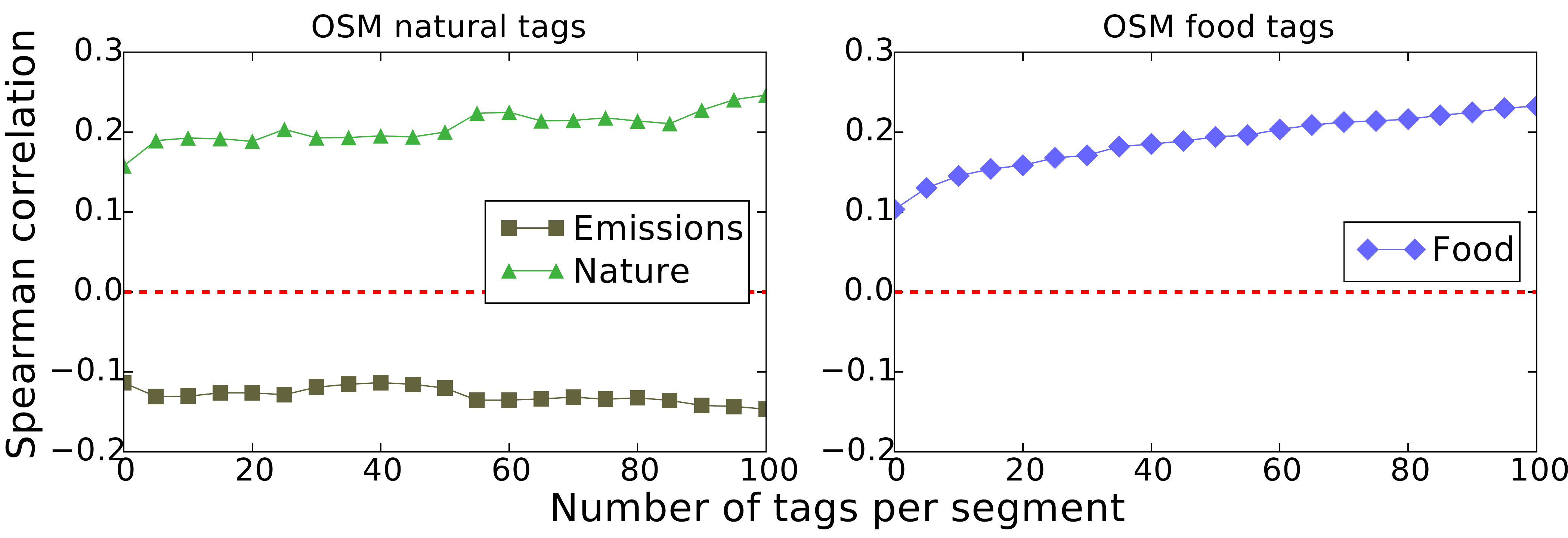}
\caption{Correlation between a street segment $l$'s smell category $S$ ($f_{S@l}$)  and the number of OSM venues as the number of picture tags per street segment increases.}
\label{fig:osm_corr_vsN}
\end{center}
\end{figure}

Then, for further validation, we also collect data about presence of natural elements and eating places from the Open Street Map (OSM) database\footnote{\url{http://wiki.openstreetmap.org/wiki/Category:Keys}}. These descriptors  identify  urban elements that are likely to be associated with certain smells. For example, the OSM \textit{natural} venue marker is used to identify natural land features;  the \textit{vegetation} and  \textit{surface} venues include natural elements such as tree, wood, and grassland; and  the \textit{cuisine} marker is associated with venues that serve food of any kind, mostly restaurants and markets. Having those OSM venues at hand, we simply count them and correlate the counts with our 10-category description of smell. As one would expect, the presence of nature and emission smells correlates  with OSM natural venues (positively and negatively, respectively), while food smells correlate with OSM cuisine venues (Figure~\ref{fig:osm_corr_vsN}).

\section{The when and where of smell} \label{sec:time}

To increase tourism, cities have been encouraged to remove negatively perceived odors and introduce more pleasant varieties~\cite{dann02}. To spot those varieties, one needs to know \emph{when} and \emph{where} to smell them.  We turn to the question of when first.

\begin{figure}[t!]
		\begin{center}
		\includegraphics[width=0.99\columnwidth]{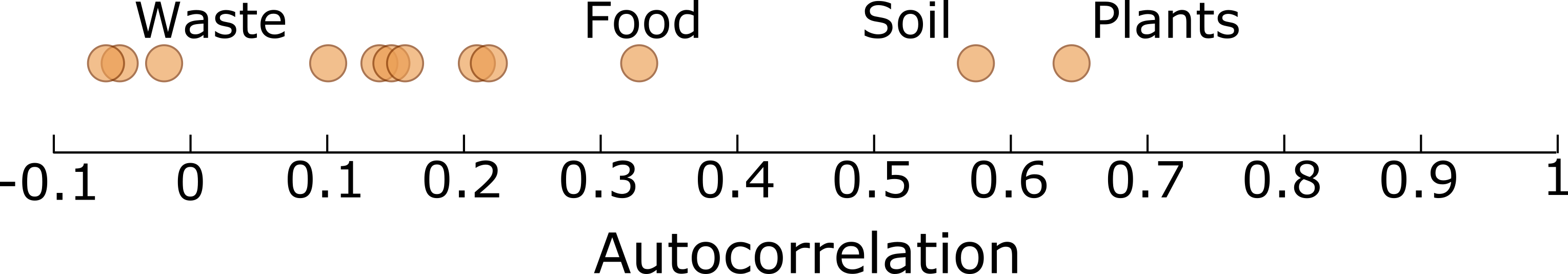}
		\caption{Temporal autocorrelation $R_{S}$ with a 12-months time lag (seasonality) of smell category $S$. Smells with autocorrelation close to 0 are unpredictable (non seasonal), while those close to 1 (e.g., plant smells) are predictable (seasonal).}
		\label{fig:autocorrelation}
		\end{center}
\end{figure}

\mbox{ } \\
\textbf{Smell of the season.} One way of determining seasonal smells is to determine the smells that show predictable temporal patterns during an entire year. To this end, we consider the time series of the relative frequency of a smell in each month, and do so for 10 years. These time series represent stationary processes that exhibit different degrees of seasonality. A standard method to measure the periodicity of this type of series is the \textit{autocorrelation} $R$. This is  the cross-correlation of the temporal signal with itself considering a fixed time window $\tau$. Specifically, we compute the autocorrelation as follows:
\begin{equation}
R_{S,\tau} = \frac{E\left[ (f_{S,t} - \mu_S) \cdot (f_{S,t+\tau} - \mu_S)  \right]}{\sigma_S^2}
\label{eq:autocorrelation}
\end{equation}
where $E$ is the expected value, $f_{S,t}$ is the fraction of the smell category $S$ at month $t$, $\tau$ is the time window we set at 12 months (as we are interested in yearly seasonality), and $\mu_S$ and $\sigma_S^2$ are the average and variance of the whole time series for the smell category $S$. $R$ ranges in $[-1,1]$ (Figure~\ref{fig:autocorrelation}). We find  that the most predictable (seasonal) smells are those of plants ($R=0.64$) followed by soil ($R=0.58$) and food ($R=0.32$). All other smells are non-seasonal, the one closest to 0 being the smell of waste ($R=-0.02$).

 \begin{table}
				\begin{center}
        \footnotesize
				\scalebox{0.9}{
				\begin{tabular}{c|cc|cc}
             & \multicolumn{2}{c|}{\textit{London}} & \multicolumn{2}{c}{\textit{Barcelona}} \\
						\hline
						\textbf{Month}	&	\textbf{Smell}	&	\textbf{Where}	&	\textbf{Smell}	&	\textbf{Where} \\
						\hline
						Jan	&	Soil	&	St. James' Park &	Soil	&	Parc Guell \\
						Feb	&	Soil	&	St. James' Park &	Soil	&	Parc Guell \\
						Mar	&	Soil	&	St. James' Park	&	Plants	&	 Parc Ciutadella\\
						Apr	&	Plants	&	St. James' park	&	Food	&	Boqueria \\
						May	&	Plants	&	Ranelagh Gardens &	Plants	& Parc Guell \\
						Jun	&	Plants	&	Regent's Park	&	Traffic	&	Pl. de Espanya \\
						Jul	&	Plants	&	Regent's Park	&	Plants	&	Parc Ciutadella \\
						Aug	&	Traffic	&	Piccadilly Circus	&	Food	&	Boqueria \\
						Sep	&	Food	&	Central London	&	Food	&	Boqueria \\
						Oct	&	Traffic	&	Regent's Street	&	Food	&	Boqueria \\
						Nov	&	Soil	&	Russel Square	&	Food	&	Boqueria \\
						Dec	&	Soil	&	Trafalgar Square &	Food	&	Boqueria \\

            \hline
        \end{tabular}}
				\caption{The smell of the month, and where to experience it best.}
				\label{fig:temporalsmells}
				\end{center}
    \end{table}
    
    \begin{table}
 \footnotesize
\begin{tabular}{|L{8cm}|}
\hline
\textbf{Pleasant smells} \\
bread, baked, baked goods, coffee, coffees, aftershave, cut grass, grass, grassy, floral, flower, flowers, flowershop, flowery, lavender, lilies, lily, magnolia, rose, rosey, tulip, tulips, violet, violets, baby, babies, child, children, sea, seaside, countryside, cedar, cedarwood, conifer, dry grass, earth, earthy, eucalyptus, ground, leafy, leaves, old wood, pine, sandalwood, soil, tree, trees, wood, woodlands, woody, petrol, diesel, fuel, gasoline, soap powder, soap \\
\mbox{ } \\ 
\textbf{Unpleasant smells} \\

flatulence, fart, vomit, dog shit, dogshit, excrement, faeces, fart, farts, feces, manure, shit, cigarette smoke, cigarette, cigarettes, cigar, cigars, smoker, tabacco, tobacco, pee, piss, ammonia, urine, public toilet, public toilets, toilet, toilets, urinal, urinals, gone-off milk, fish, rotten fish, rotten food, rotten, rotten fruit, rotten fruits, putrid, bus, buses, car, cars, exhaust, traffic, fume, fumes, body odour, body odor, sweat, sweaty, dirty clothes \\
\hline
\end{tabular}
\caption{List of smells people generally find pleasant and those they find unpleasant.}
\label{fig:goodbadmells}
 \end{table}

\begin{figure*}[t!]
\subfloat[London \label{fig:maps_posneg_london}]{%
\includegraphics[width = 0.49\textwidth]{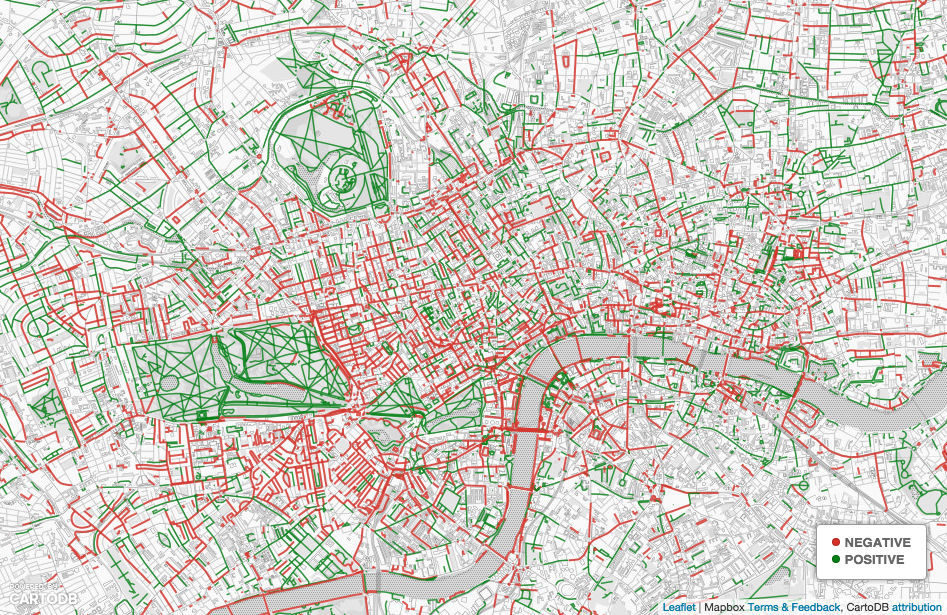}
}
\hfill
\subfloat[Barcelona \label{fig:maps_posneg_barcelona}]{%
 \includegraphics[width = 0.49\textwidth]{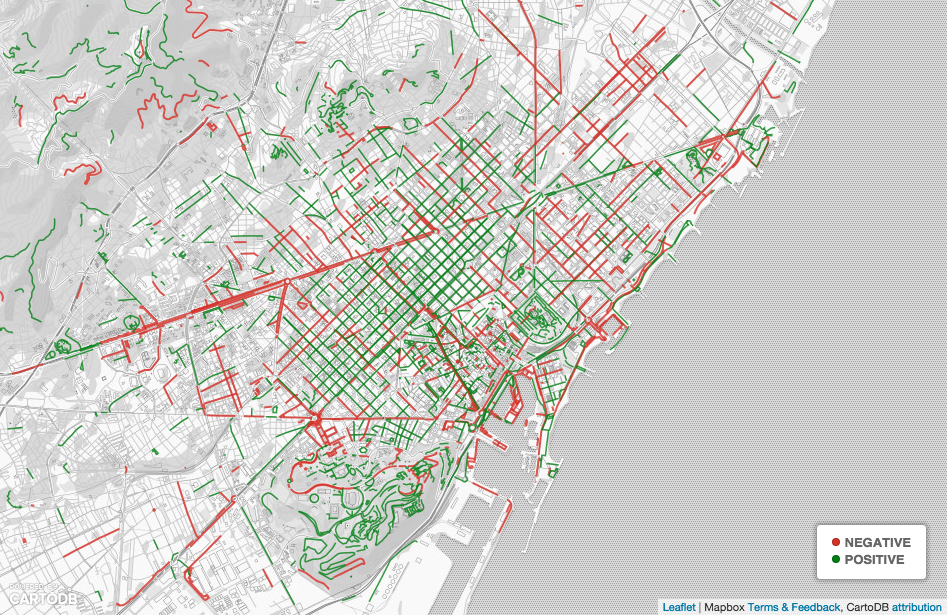}\\
	}
\caption{Maps of the olfactory pleasantness of street segments in London and Barcelona. They report the $z$-scores of pleasant smells $z_{pleasure@l}$ at each street segment $l$. The color of a street goes from green (very pleasant)  to red (very unpleasant). Pleasantness scores can be further explored at~\url{http://goodcitylife.org/smellymaps} by selecting a street segment of interest.}

\label{fig:maps_posneg_london_barcelona}
\end{figure*}

\mbox{ } \\
\textbf{Smell of the month.}   Some smells characterize not only an entire season but also specific months. To show that, for each month $t$, we compute the most frequent smell categories, that is, we rank each smell category $S$ by $f_{S,t}$ and choose the one at the top:
\begin{equation}
f_{S,t}= \frac{\textrm{\#tags in smell category } S \textrm{at month } t}{\textrm{\#tags in any smell category at month } t}
\label{eqn:fraction-smell-month}
\end{equation}

\begin{figure}[t!]
\begin{center}
\includegraphics[width=0.75\columnwidth]{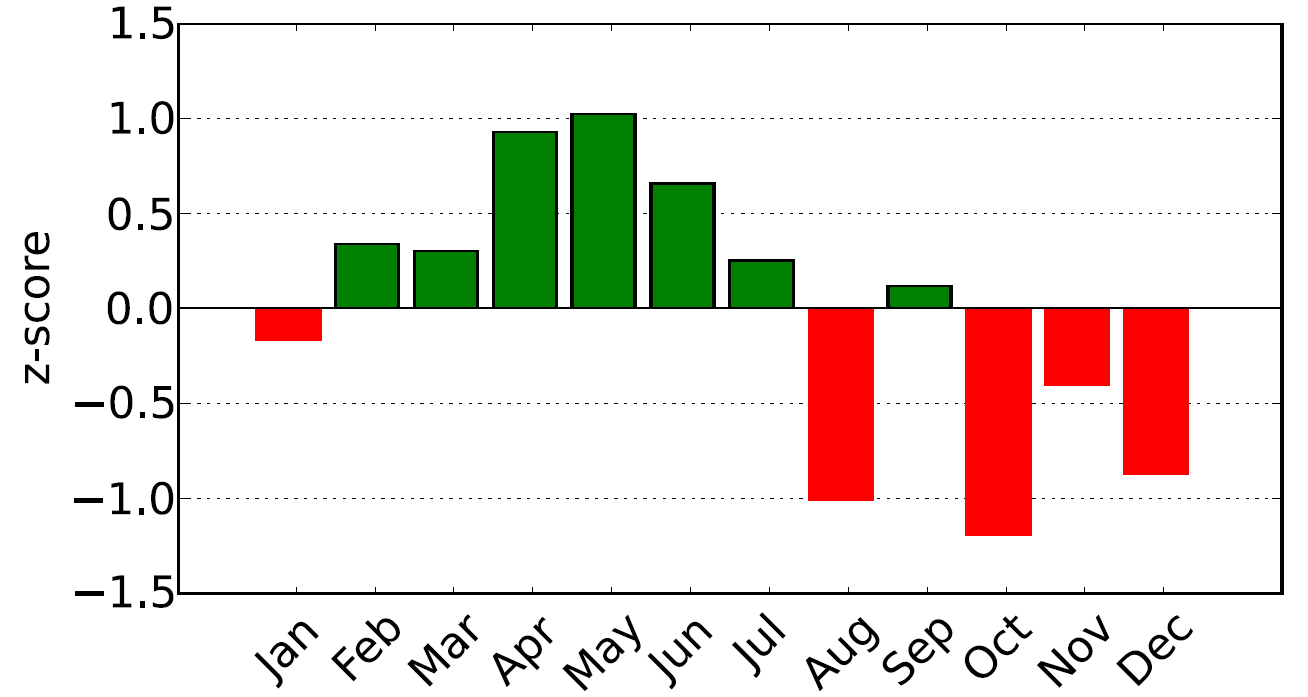}
		\caption{Olfactory pleasantness $z_{pleasure,t}$ of each month based on 10-year statistics.}
		\label{fig:temporal_london}
		\end{center}
\end{figure}

The results are reported in Table~\ref{fig:temporalsmells}: the most frequent categories include trees \& soil from November to March, and flowers \& plants from April to July. August and September deviate from the pattern and are characterized by traffic (August) and food (September).  Plants can be smelled in Ranelagh Gardens (during the Flower Festival in May) and Regent's Park. However, not all months are equal: there might be months that are olfactory more distinctive than others. To identify them, we compute the Shannon entropy from the vector of the smell frequencies $ <f_{S,t}>_S $ for each month $t$. We find that the least distinctive month is January, while the most distinctive ones are March, April, and May. Finally, there might be months that are more olfactory pleasant than others. To determine what is pleasant and what is not, we resort to the literature~\cite{victoria2013}, which lists the smells people tend to like and those they tend to dislike (Table~\ref{fig:goodbadmells}). To then derive a pleasantness score out of pictures for each month $t$,  we compute the \emph{pleasure} score, which is the $z$-score of the fraction of pleasant tags minus the $z$-score for the unpleasant ones:
\begin{equation}
z_{pleasure,t} = \frac{f_{pleasant,t} - \mu_{pleasant}}{\sigma_{pleasant}} - \frac{f_{unpleasant,t} - \mu_{unpleasant}}{\sigma_{unpleasant}}
\label{eqn:pleasantness}
\end{equation}
Where $\mu_{pleasant|unpleasant}$ and $\sigma_{pleasant|unpleasant}$ are the average and standard deviation of the fraction of tags reflecting pleasant (unpleasant) smells across all locations and all calendar months (considering a 10-years statistics). The higher the pleasantness score, the higher the concentration of  pleasant smells over unpleasant ones. Pleasantness is zero when the number of pleasant tags and that of unpleasant ones are both equal to their average values. We plot the pleasantness $z$-scores for twelve months (Figure~\ref{fig:temporal_london}), and find that the peak is registered in May, which turns out to be the best smelling month of the year. 

\mbox{ } \\
\textbf{The where of smell.} Knowing the smells of the month, we now turn to determine where to perceive them. We find that, in London, soil smells are best experienced near Trafalgar Square,  Russel Square, and St James' Park (Table~\ref{fig:temporalsmells}). We find similar results for Barcelona, with a greater emphasis for food smells though. These are best experienced in La Boqueria, which is the main food market in the center of the city.  We then identify the most (un)pleasant locations (street segments) by computing $z_{pleasure@l}$ for each location $l$:
\begin{equation}
z_{pleasure@l} = \frac{f_{pleasant@l} - \mu_{pleasant}}{\sigma_{pleasant}} - \frac{f_{unpleasant@l} - \mu_{unpleasant}}{\sigma_{unpleasant}}
\label{eqn:pleasure@l}
\end{equation}
We then map those values (Figure~\ref{fig:maps_posneg_london_barcelona}) and see that parks tend to have the most pleasurable smells, while main roads are infested with the least pleasurable.

\section{The emotion of smell} \label{sec:emotions}

Looking at a location through the lens of social media makes it possible to study that location from different points of views.  So far we have studied the spatio-temporal dynamics of smell. Yet, the sense of smell has a highly celebrated link with other aspects, most notably with emotions. The nose has direct access to the amygdala, the part of the brain that controls emotional response~\cite{gilbert2008nose}. As such, smells have a considerable effect on our feelings and our behavior. 

Therefore, we set out to study the relationship between the smellscape and emotions on our data. To do so, we need to have a lexicon of emotion words. We use two of them: the ``Linguistic Inquiry Word Count'' (LIWC)~\cite{pennebaker2013secret}, that classifies words into positive and negative emotions, and the ``EmoLex'' word-emotion lexicon~\cite{mohammad2013crowdsourcing}, that classifies words into eight primary emotions based on Plutchik's psycho-evolutionary theory~\cite{plutchik1991emotions} (i.e., \textit{anger}, \textit{fear}, \textit{anticipation}, \textit{trust}, \textit{surprise}, \textit{sadness}, \textit{joy}, and \textit{disgust}).

\begin{figure}[t!]
\begin{center}
\includegraphics[width=0.99\columnwidth]{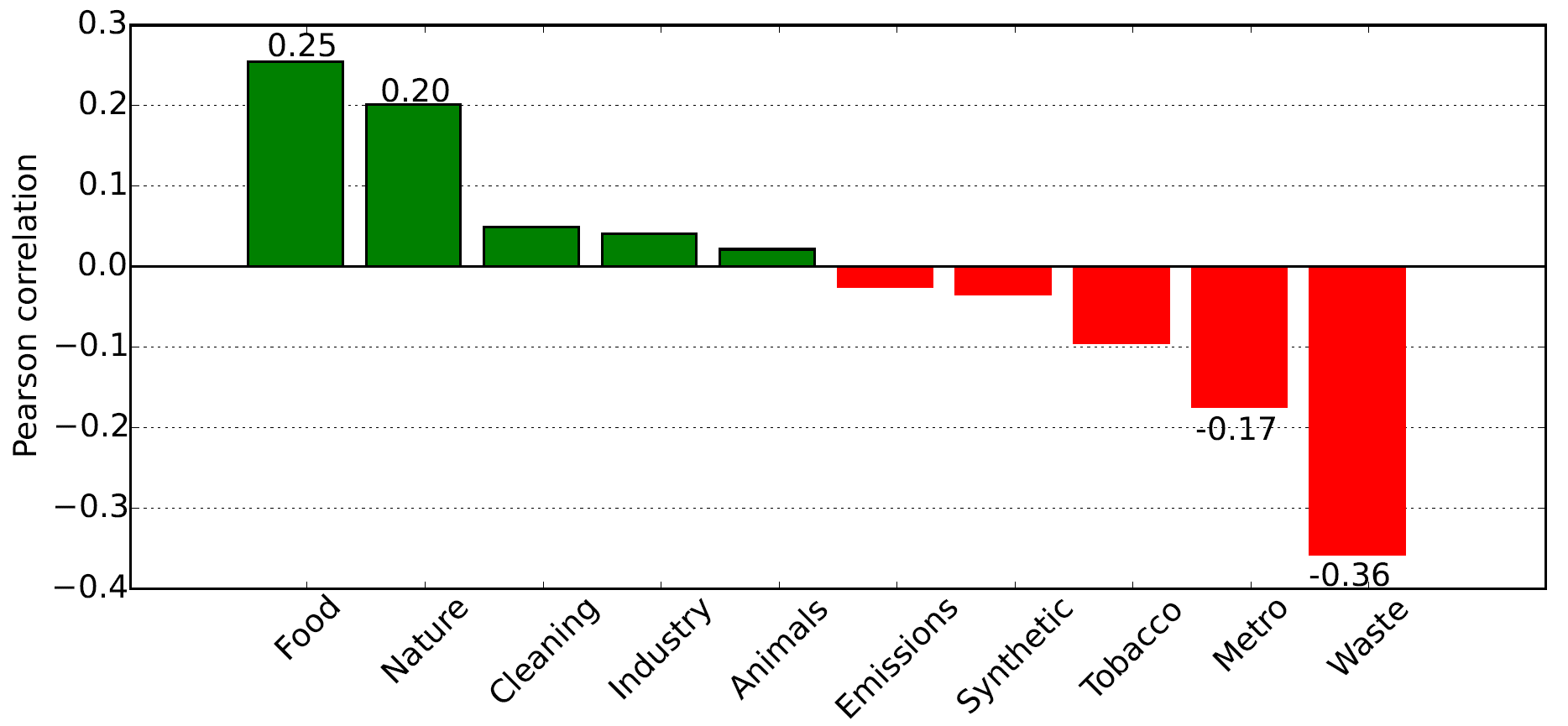}
\caption{Pearson correlation between presence of smell category $f_{S@l}$ and presence of positive emotion words $z_{sentiment@l}$. Values are computed considering street segments with at least 30 smell tags. The $p$-value is always $<$ $0.01$.}
\label{fig:liwc_london}
\end{center}
\end{figure}

\begin{figure}[t!]
\begin{center}
\includegraphics[width=0.99\columnwidth]{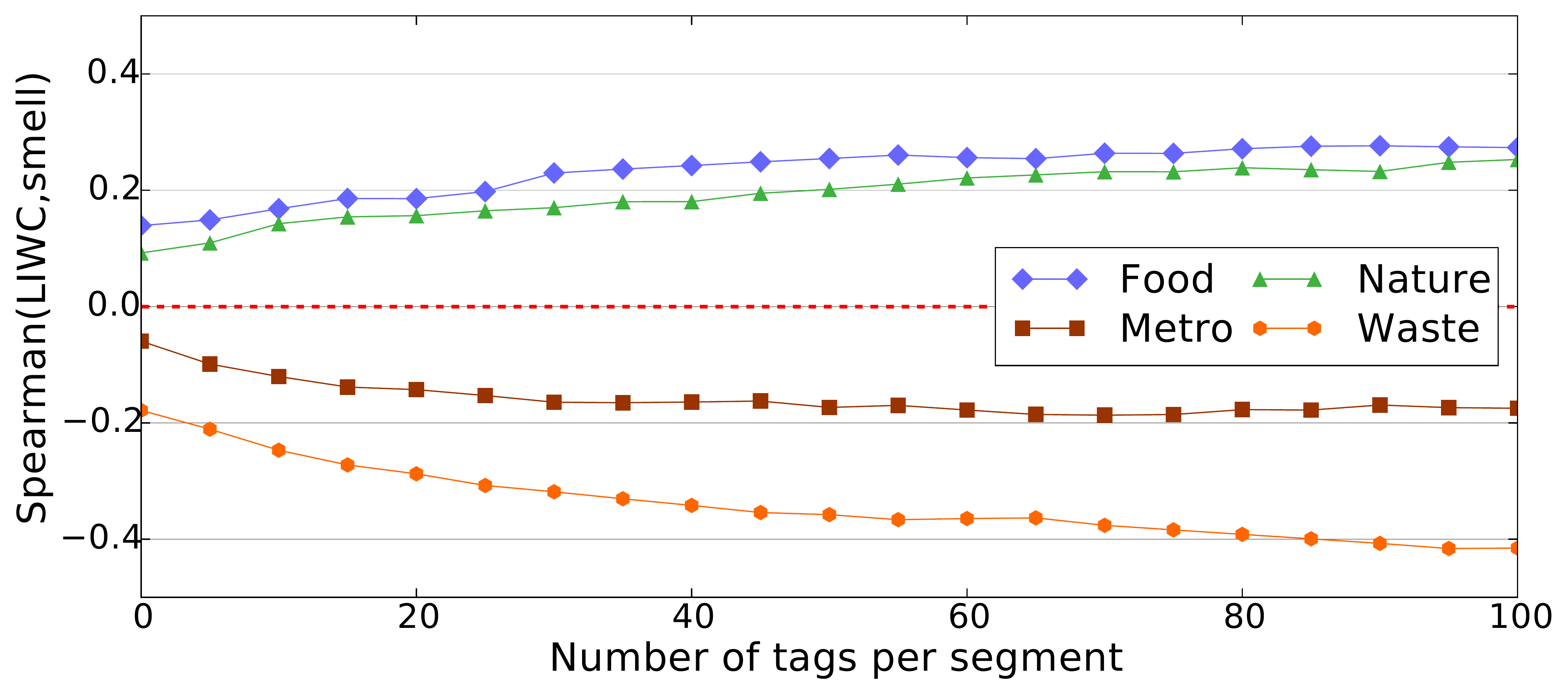}
\caption{Spearman correlation between presence of smell category $f_{S@l}$ and sentiment score $z_{sentiment@l}$ as the number of tags per street segment increases. The four smell categories with the strongest correlations are shown. The $p$-value (across all street segments) is always $<$ $0.01$.}
\label{fig:liwc_smell_corr_vsN}
\end{center}
\end{figure}

From those two lexicons, we use the polarity of positive and negative words to compute the \emph{sentiment score} by subtracting the $z$-score of the negative tags from the $z$-score of the positive ones:
\begin{equation}
z_{sentiment@l} = \frac{f_{positive@l} - \mu_{positive}}{\sigma_{positive}} - \frac{f_{negative@l} - \mu_{negative}}{\sigma_{negative}}
\label{eqn:sentiment@l}
\end{equation}

From EmoLex, we compute the $z$-score of its eight individual emotions:
\begin{equation}
f_{E@l}= \frac{\textrm{\#tags emotion category } E \textrm{ at location } l}{\textrm{\#tags in any emotion category at location } l}
\label{eqn:fraction-emotion}
\end{equation}

\begin{figure*}[t!]
\begin{center}
\includegraphics[width=0.95\textwidth]{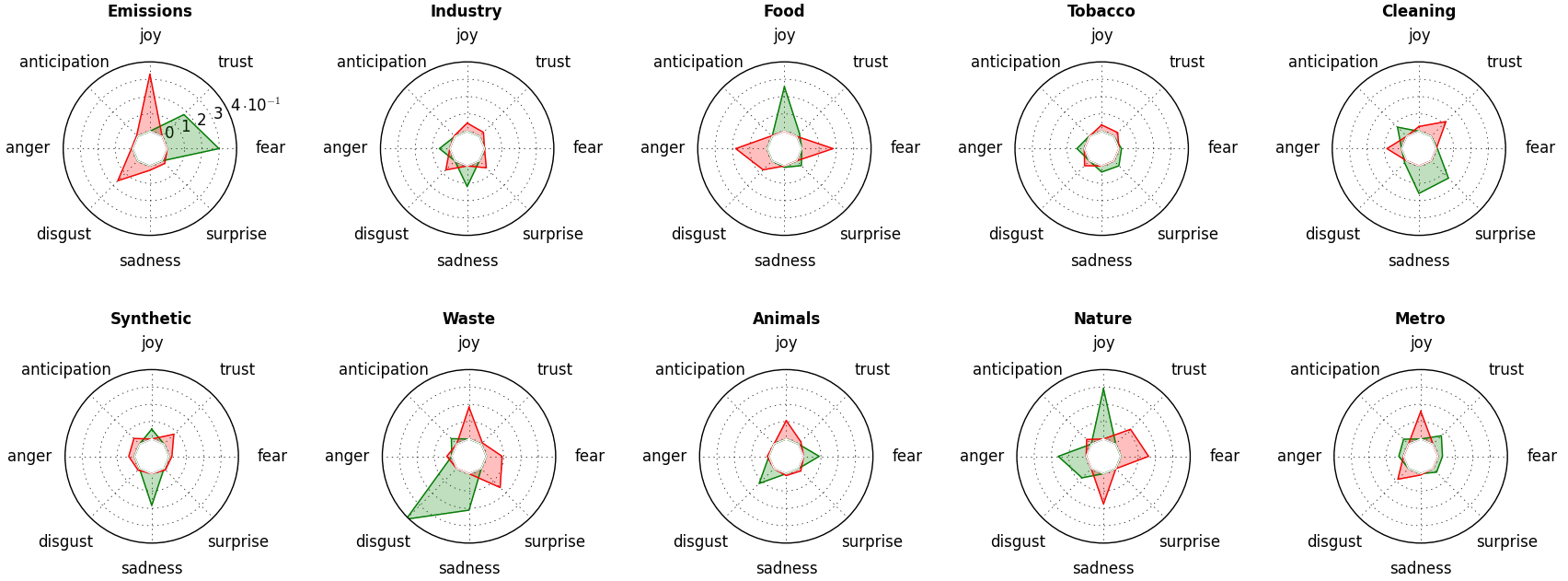}
\caption{Correlation between the fraction $f_{S@l}$ of tags in each smell category $S$ and the fraction $f_{E@l}$ of tags in each Plutchik's emotion category $E$. Positive correlations are in green, and negative ones are in red. All correlations are statistically significant at the level of $p$~$<$~$0.01$.}
\label{fig:emolex}
\end{center}
\end{figure*}

\mbox{ } \\ 
\textbf{Emotions and smells.} After computing the Pearson correlations between the presence of each smell category $S$ and LIWC sentiment scores (Figure~\ref{fig:liwc_london}), we find that positive sentiment tags are found in streets with food and nature smells, while negative sentiment tags are found in streets with waste and metro smells. To be significant, those correlations do not require a large number of tags per street: it depends on the smell categories but 50 tags are usually enough (Figure~\ref{fig:liwc_smell_corr_vsN}). The same results are obtained when the sentiment score is computed from  the classification of \emph{EmoLex}. Furthermore, this latter lexicon allows for studying finer-grained emotions as it considers eight emotional constructs.  We correlate the fraction $f_{S@l}$ of tags in smell category $S$ at each location  and the fraction of tags in each emotion category $E$ and show the results in Figure~\ref{fig:emolex}. We observe that waste correlates positively with disgust and sadness but negatively with joy. Similarly, where emission-related smells are present, joy-related words are absent. Interestingly, emission smells positively correlate with trust and fear (whose combination is interpreted as \textit{submission} by Plutchik's theory). Those correlations are especially strong in dense urban areas in which the fear generated by intense traffic likely mixes with the trust and safety generated by the presence of crowds~\cite{speck2012walkability}. 

\begin{figure}[t!]
\begin{center}
\includegraphics[width=0.99\columnwidth]{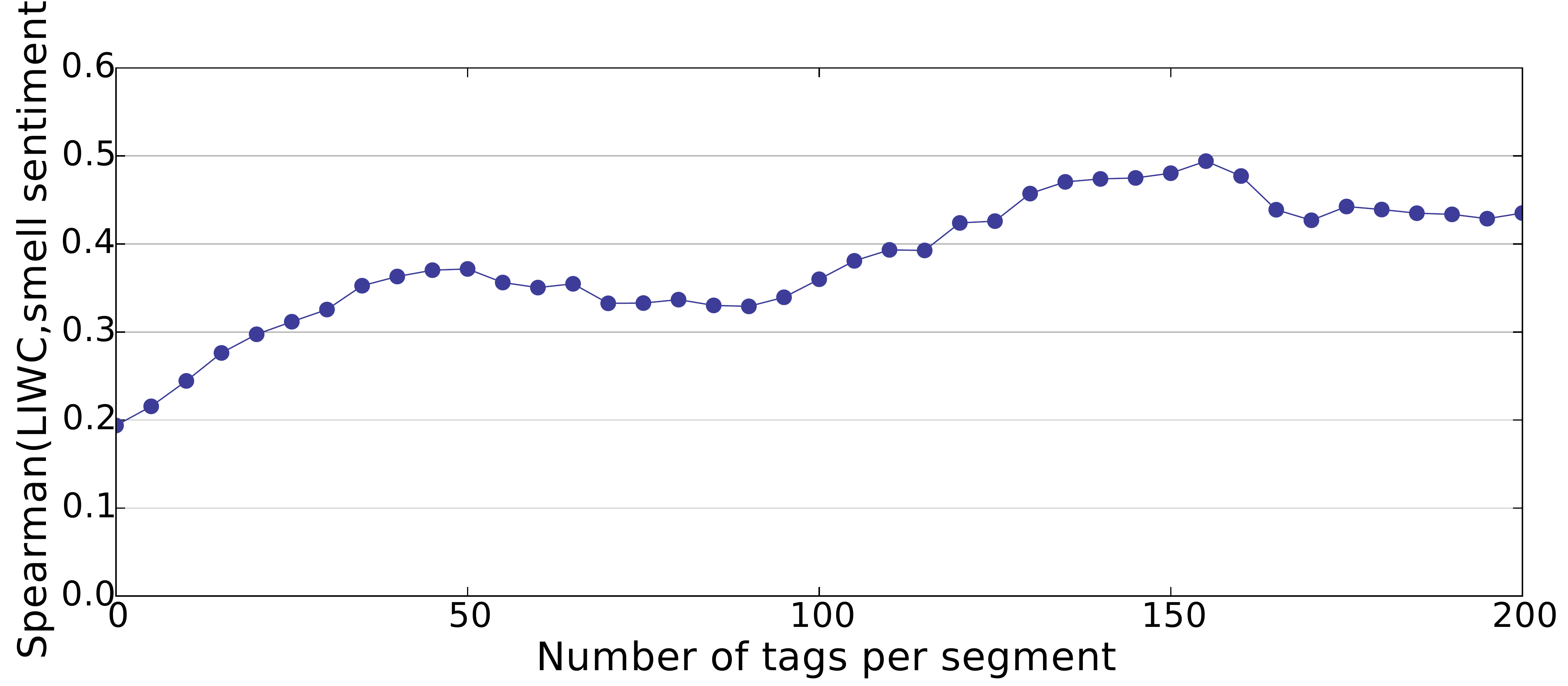}
\caption{Spearman correlation between sentiment score $z_{sentiment@l}$ and presence of pleasant smell (tags) $z_{pleasure@l}$ as the number of  tags per street segment ($x$-axis) increases. The $p$-value is always $<$ $0.01$}
\label{fig:liwc_posneg_corr}
\end{center}
\end{figure}

\mbox{ } \\ 
\textbf{Emotions and pleasant smells.} In Section~\ref{sec:time}, to find the most olfactory pleasant month, 
we have been able to distinguish pleasant from unpleasant smells based on the literature (Table~\ref{fig:goodbadmells}). One could reasonably hypothesize that areas with (un)pleasant smells are characterized by specific emotions. To verify that, we compute the Pearson correlation between a street segment's (un)pleasant smells (as per Formula~\ref{eqn:pleasantness} in Section~\ref{sec:time}) and the segment's sentiment. We find that, indeed, locations with pleasant smells tend to be associated with positive emotion tags (with correlation $r$ up to 0.50), while locations with unpleasant smells tend to be associated with negative ones. The correlations change depending on the number of tags per street segment but are quite stable after 150 tags (Figure~\ref{fig:liwc_posneg_corr}).

\section{The color of smell}
\label{sec:colors}

We have seen that there is an association between emotional and olfactory layers. Emotions are triggered not only by the sense of smell but also by the sense of sight. What we see impacts how we feel. When detecting images, our retina generates nerve impulses for varying colors. In its most basic form, our vision revolves around colors and, consequently, early studies have related colors to emotions. 

In marketing, colors are widely used to  influence consumers' emotions and perceptions. Despite cross-cultural differences, there are significant cross-cultural similarities regarding which emotional states people associate with different colors. For example, the color red is often perceived as strong and active~\cite{widermann11}, tones of black lead to feelings of grief and fear, while green tones are often associated with good taste~\cite{aslam06}. 

\begin{figure}[t!]
\includegraphics[width=0.32\columnwidth]{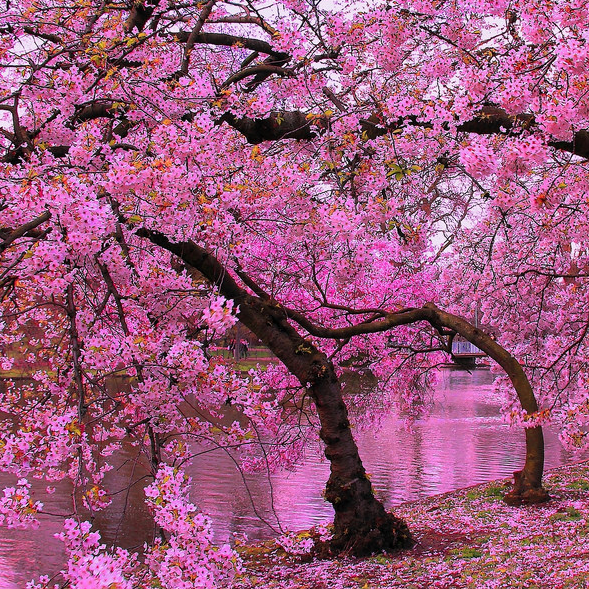} 
\includegraphics[width=0.32\columnwidth]{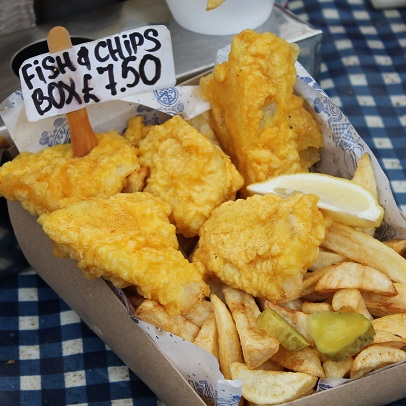} 
\includegraphics[width=0.32\columnwidth]{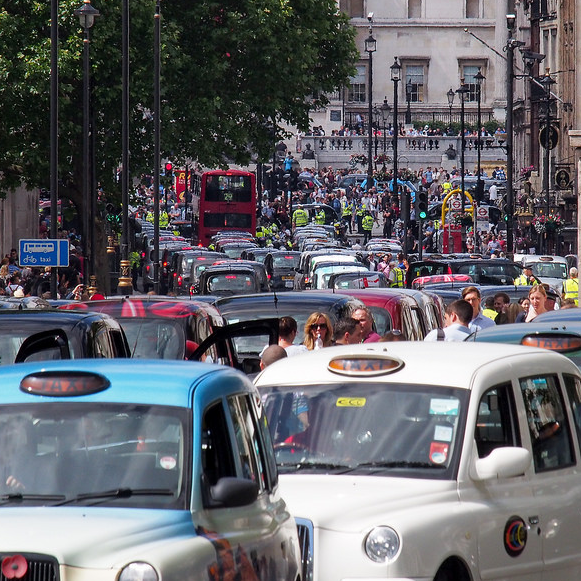}
\\
\includegraphics[width=0.32\columnwidth]{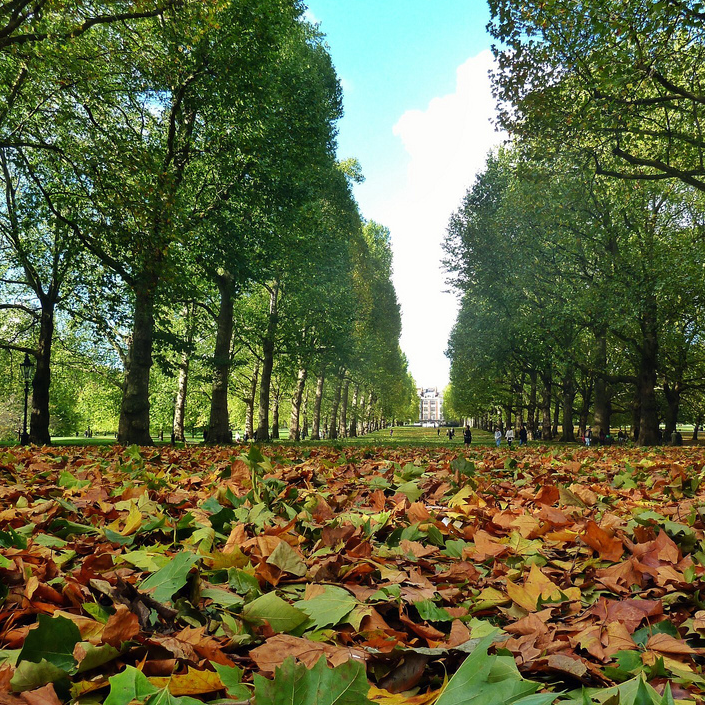} 
\includegraphics[width=0.32\columnwidth]{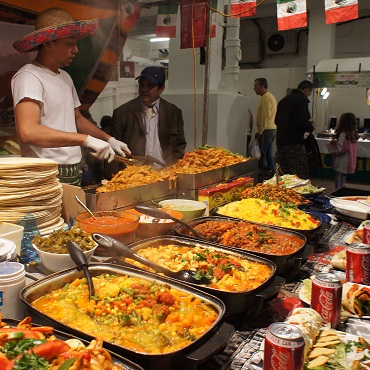} 
\includegraphics[width=0.32\columnwidth]{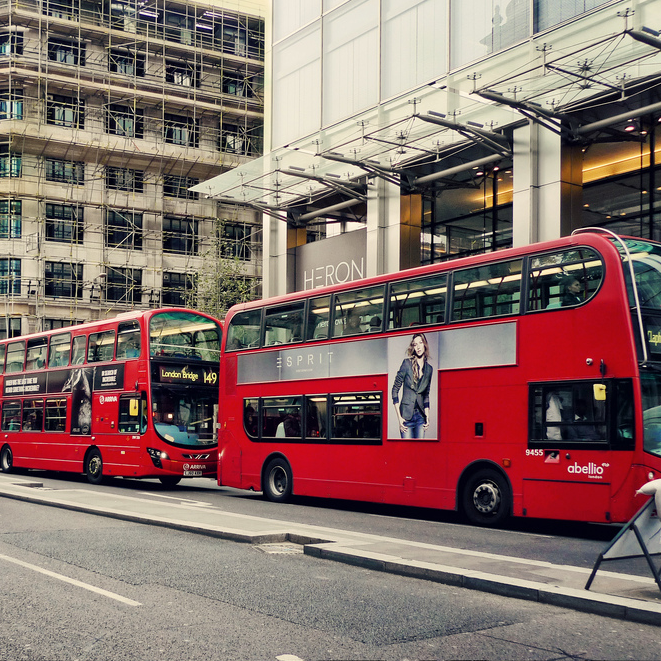}
\caption{Examples of pictures in which a smell tag and a color tag co-occur. The first row has (from left to right) tree+violet, food+brown, and traffic+gray. The second row has tree+green/brown, food+brown, and traffic+red. }
\label{fig:examples}
\end{figure}

We therefore explore this last relationship: that between smellscape and colors. As a first step, we need to extract colors from our pictures\footnote{We discard black and white photos.} (Figure~\ref{fig:examples}). We do so not from the images themselves (to avoid the noise introduced by the images with multiple colors) but from tags. To textually extract colors from tags, we build a color term dictionary by grouping 249 color nuances into ten main colors (collating the colors into fewer chromatic categories greatly reduces  spurious matches)\footnote{\url{www.farb-tabelle.de/en/table-of-color.htm}}: black, blue, yellow, gray, green, orange, red, violet, white, and yellow. 

Technically, having color $c$ at hand, we compute the strength of its association with smell $s$ based on color-smell co-occurrences, normalized by the total for that color:
\begin{equation}
strength_{s,c} = \frac{\frac{p_{cs}}{p_c+p_s}}{\sum_{c}{\frac{p_{cs}}{p_c+p_s}}}
\end{equation}

where  $p_{cs}$ is the number of photos in which $c$ and $s$ co-occur,  $p_{c}$ is the number of photos associated with color $c$, and $p_{s}$ is the number of photos associated with smell $s$. To make  strength scores comparable across smell categories, we divide the ratio by the rescaling factor $\sum_{c}{\frac{p_{cs}}{p_c+p_s}}$.  In computing the strength score, we make sure to consider only the (smell,color) associations for which we have at least 10 pictures  (i.e., $p_{cs} \geq 10$), effectively avoiding  spurious associations.

\begin{figure}[t!]
\begin{center}
\includegraphics[width=0.68\columnwidth]{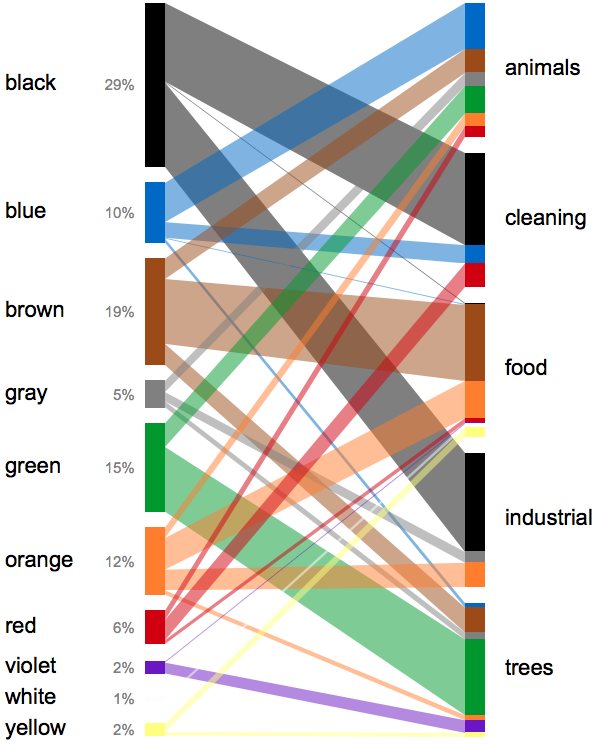}
\caption{Bipartite graph of smell-color associations (best seen in color). The ticker a line, the higher the association strength $strength_{S,c}$ between smell category $S$ and color $c$. The percentage next to each color is the average fraction of that color across all smells. For example, black characterizes any smell 29\% of the times. Only the five smells with the lowest entropy of color mixture are shown. An interactive version is available at \url{http://goodcitylife.org/smellymaps/chromosmell}.}
\label{fig:graph}
\end{center}
\end{figure}

\begin{figure}[t!]
\begin{center}
\includegraphics[width=0.9\columnwidth]{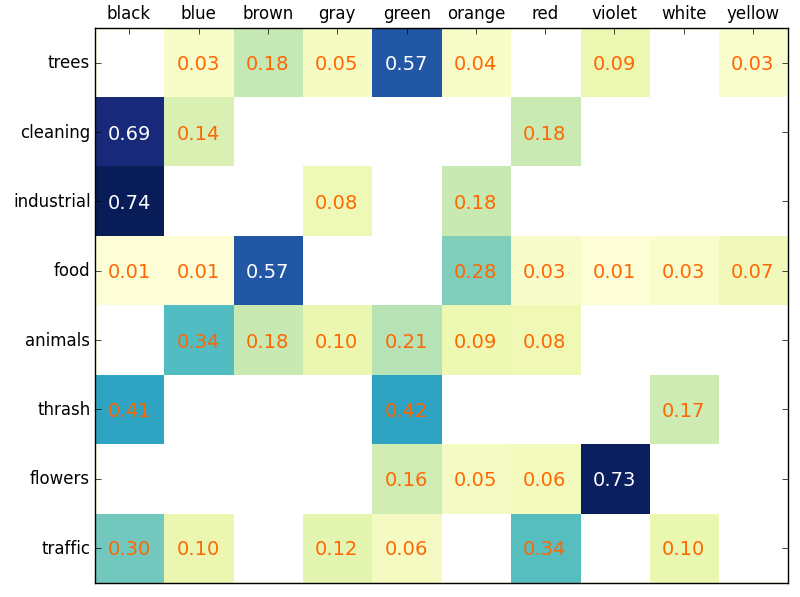}
\caption{Matrix of smell-color associations. The darker, the stronger the association. Each number is $strength_{S,c}$, which reflects the association between color $c$ and  smell category $S$. Smells are sorted in ascending order of entropy of color mixture (entropy has nothing to do with association strength and is the Shannon entropy of each row in the matrix). Smells at top are associated with very few colors (e.g., trees are predominantly associated with green), while smells at the bottom are associated with a variety of colors (e.g., flowers are colorful).}
\label{fig:colorsmell_matrix}
\end{center}
\end{figure}

Instead of determining the color of each individual smell (which would suffer from data sparsity), it would be more informative to determine the color of an entire smell category. To do so, we aggregate the fine-grained color-smell associations. A smell category consists of individual smell words. For each smell word $s$ (e.g., violet),  we determine the most representative color by selecting color $c$  ($c_{strongest}$) with the highest $strength_{s,c}$ (e.g, purple). To then determine the color of the smell category $S$ (e.g., nature), we sum the strengths for all colors in that category: $strength_{S,c}$= $\sum_{\forall s \in S} strength_{s,c_{strongest}}$. We do so by considering only the (smell,color) associations for which we have at least 10 pictures. The color associations (Figure~\ref{fig:graph}) meet expectations: traffic is associated with black and red (likely from road, smoke, traffic lights and red buses); industrial smells with black as well; trees and soil with green; flowers with green, violet, and orange; and food with brown and orange. To allow for reproducibility  and the re-use of our results, we spell out all the associations strengths in Figure~\ref{fig:colorsmell_matrix}.

We are not the first to relate colors to smell. It has been repeatedly shown that there is a reliable multi-sensory interaction between odors and colors (e.g., between the color yellow and the odor of bergamot). By asking study participants to associate those two things, researchers have found consistent associations~\cite{gilber96}. To go beyond explicit matches, Dematte \emph{et al.} used an indirect association measure (called implicit association test~\cite{greenwald98}) and  showed that color-odor associations are both systematic and robust~\cite{demate06}. All those studies have focused on associations made by individuals; we have studied the same associations at \textit{geographic} level. We have done so because our insights can inform the design of smell-related technologies. As a short-term example, consider mapping. When showing a specific smell on the city map or on any user interface, it is important to know which color to use. As a long-term example, consider virtual reality. A multi-sensory virtual reality experience should effectively match the visual (including colors) and the olfactory experience. In fact, it has been shown that, to be effective, smell must match context (studies refer to that as the ``congruency'' problem~\cite{gilbert2008nose}). To see why, consider the researchers who explored the combined effect of smell and music in a gift shop~\cite{Mattila01}: consumer satisfaction increased when the store had (low-arousal) lavender and relaxing music or (high-arousal) grape-fruit and energizing music; by contrast, no effect was registered when there was a mismatch between music and smell.

\section{Discussion}
\label{sec:discussion}

Before concluding, we briefly discuss our work's limitations, theoretical implications, and practical implications. 

\mbox{} \\
\textbf{Limitations.} This work has considered the ``average'' urban smellscape. Yet, smell is mediated by individual factors (e.g., age, gender), geographic factors (e.g., climatic conditions), and contextual factors (e.g., city layout)~\cite{victoria2013}. Also, not all odors are the same. It is difficult to capture fleeting odors from social media, not least because those odors are localized in space and time. Our analysis, instead, is able to capture what olfactory researchers call base smell notes (uniformly distributed across the city) and mid-level notes (localized in specific areas of the city).

\mbox{} \\
\textbf{Theoretical Implications.} This work has shown that social media could be used to capture insights about urban smellscapes both spatially and temporally. Olfactory researchers have so far focused on negative characteristics of smell. Now, based on unobtrusive data capturing from social media users, the same researchers could move forward and study the positive role that odors can play in the environmental urban experience. We have also shown for the first time how sensorial perceptions can be mapped to orthogonal dimensions like the ones of emotions and colors.

All these aspects are not incremental. On the contrary, they pertain to totally different domains than that of smell. Think about a park. As for smell, it falls into the nature category. As for emotions, instead, it is multi-faceted: different parts of the park speak to opposite emotions (e.g., lavender is calming, and fresh grass is energizing). Then, as for time, the park drastically changes across seasons. Finally, as for colors, the chromatic reality implicitly suggested by the pictures goes beyond the obvious description of the stereotypical green park.

\mbox{} \\
\textbf{Practical Implications.} Our work might help a variety of stakeholders. \emph{Urban planners} could go beyond the attempt of mitigating bad odors and their potential negative impact and could use, instead, social media tools to monitor the whole spectrum of emotion associated with the smellscape. This might inform policies to incentivize the development of areas with pleasurable smells. On a similar note, the tourism industry could capitalize on olfactory opportunities by facilitating the discovery and exploration of places that are not conventionally included in touristic tours. \emph{Computer scientists} have worked on way-finding tools that suggest not only shortest routes between two points in the city but also short ones that are beautiful~\cite{quercia14shortest} or visually distinctive~\cite{noulas12random,vancaneyt11time}. Now they could look into recommending routes that are olfactorily pleasant, as our work provides a principled and scalable method to capture smell. \emph{Interaction designers} could inform their work with the knowledge of how color, smell, and emotions are interrelated and, as such, know which color to use on, say, a map when graphically representing smell. As a proof-of-concept, we built an interactive map\footnote{\url{http://goodcitylife.org/smellymaps/}} that allows to navigate the smellscape of central London. The interface has been selected as one of the finalist projects in the Insight Competition by CartoDB\footnote{\url{http://blog.cartodb.com/insight-finalists}} and is exhibited  at the Storefront for Art and Architecture in New York City. As for the last stakeholder, the \emph{general public} could now nurture a critical voice for the positive and negative role that smell has to play in the city.

\section{Conclusion}
In Japan, one hundred sites have been declared as protected because of their `good fragrance'. By contrast, in the rest of the world,  environmental smells have received little attention.  Our urbanization age, however, results into closer proximity between people and activities and, as such, faces new olfactory challenges. To tackle those challenges, we need to capture the complex olfactory fragments of our cities. We have used social media to do so and studied the interplay between dynamics of different nature: spatio-temporal, emotional, and chromatic. To build upon this work,  we are currently exploring three main directions. First, we are working on a smell app to capture the fleeting odors that cannot be extracted from social media. Second, we have teamed up with epidemiologists to determine under which conditions social media  complements the use of extremely costly  air sampling devices. Third, we are designing a study to assess whether it is possible to change the behavior of city dwellers by making urban smell visible. The goal of this work has been to make visible what is normally not and, ultimately,  to make it possible to create more fulfilling, humanistic, and sustainable urban environments. 

\section*{Acknowledgments}
\small
We thank researchers at the Centre for Research in Environmental Epidemiology (CREAL) for providing air quality data for the city of Barcelona. 

\balance{}
\small
\bibliographystyle{aaai}
\bibliography{bibliography}

\end{document}